\documentclass{osa-article}

\usepackage{wasysym}
\usepackage[colorinlistoftodos]{todonotes}

\journal{oe}

\begin{document}

\newcommand{\um}{\textmu m}
\newcommand{\uA}{\textmu A}

\title{A kilopixel array of superconducting nanowire single-photon detectors}

\author{Emma E. Wollman,\authormark{1,3} Varun B. Verma,\authormark{2,4}  Adriana E. Lita,\authormark{2} William H. Farr,\authormark{1} Matthew D. Shaw,\authormark{1} Richard P. Mirin,\authormark{2} and Sae Woo Nam\authormark{2}}

\address{\authormark{1}Jet Propulsion Laboratory, California Institute of Technology, 4800 Oak Grove Dr., Pasadena, CA 91109, USA\\
\authormark{2}National Institute of Standards and Technology, 325 Broadway, Boulder, CO 80305, USA\\
\authormark{3}emma.e.wollman@jpl.caltech.edu\\
\authormark{4}varun.verma@nist.gov}

\begin{abstract*}
We present a 1024-element imaging array of superconducting nanowire single photon detectors (SNSPDs) using a $32 \times 32$ row-column multiplexing architecture. Large arrays are desirable for applications such as imaging, spectroscopy, or particle detection.
\end{abstract*}

%%%%%%%%%%%%%%%%%%%%%%%%%%  body  %%%%%%%%%%%%%%%%%%%%%%%%%%
\section{Introduction}
Superconducting nanowire single-photon detectors (SNSPDs) are among the highest-performing photon counters in terms of efficiency, speed, dark counts, and range of wavelength sensitivity. Traditionally, SNSPDs have had the greatest impact in applications that rely on their high timing resolution, including optical communication \cite{Boroson2014, Shaw2017, Biswas2018}, quantum optics \cite{Takesue2007, Chen2008, Clausen2011, Stevens2010, Bussieres2014}, and lidar \cite{McCarthy2013}. Recently, new applications seek to take advantage of SNSPDs' ultra-low dark count rates, which can be below $10^{-4}$~counts/s \cite{Hochberg2019}. In fields such as dark matter detection \cite{Hochberg2019} and space-based astronomy \cite{OST2018}, SNSPDs offer high-efficiency detection of low-flux signals, low intrinsic dark-count rates, and the possibility of gating out false events due to their timing resolution. However, these applications generally require larger arrays than are currently available, both in terms of number of elements and active area. Large SNSPD arrays could additionally be useful for quantum imaging, time-resolved imaging, or lidar applications.

The most mature and straightforward architecture for multi-pixel SNSPD arrays is direct readout, where each pixel has its own readout channel. To date, arrays of 64 pixels have been demonstrated using direct readout \cite{Miki2014, Shaw2017}. As pixel counts increase, direct readout becomes unfeasible due to the increasing heat load of high-speed cables and the limited cooling power of typical cryostats. Several architectures for cryogenic multiplexing of SNSPD arrays have been successfully demonstrated for array sizes up to 64 pixels \cite{McCaughan2018}, including row-column multiplexing \cite{Allman2015}, SFQ readout \cite{Miyajima2018}, pulse amplitude multiplexing \cite{Zhao2013}, and frequency multiplexing \cite{Doerner2017}. Additionally, a delay-line based SNSPD imager has been demonstrated with $\sim$~600 effective pixels \cite{Zhao2017}. Here, we report on the first kilopixel-scale SNSPD array, achieved using a $32 \times 32$ row-column multiplexing architecture.

\begin{figure}[h]
\centering\includegraphics{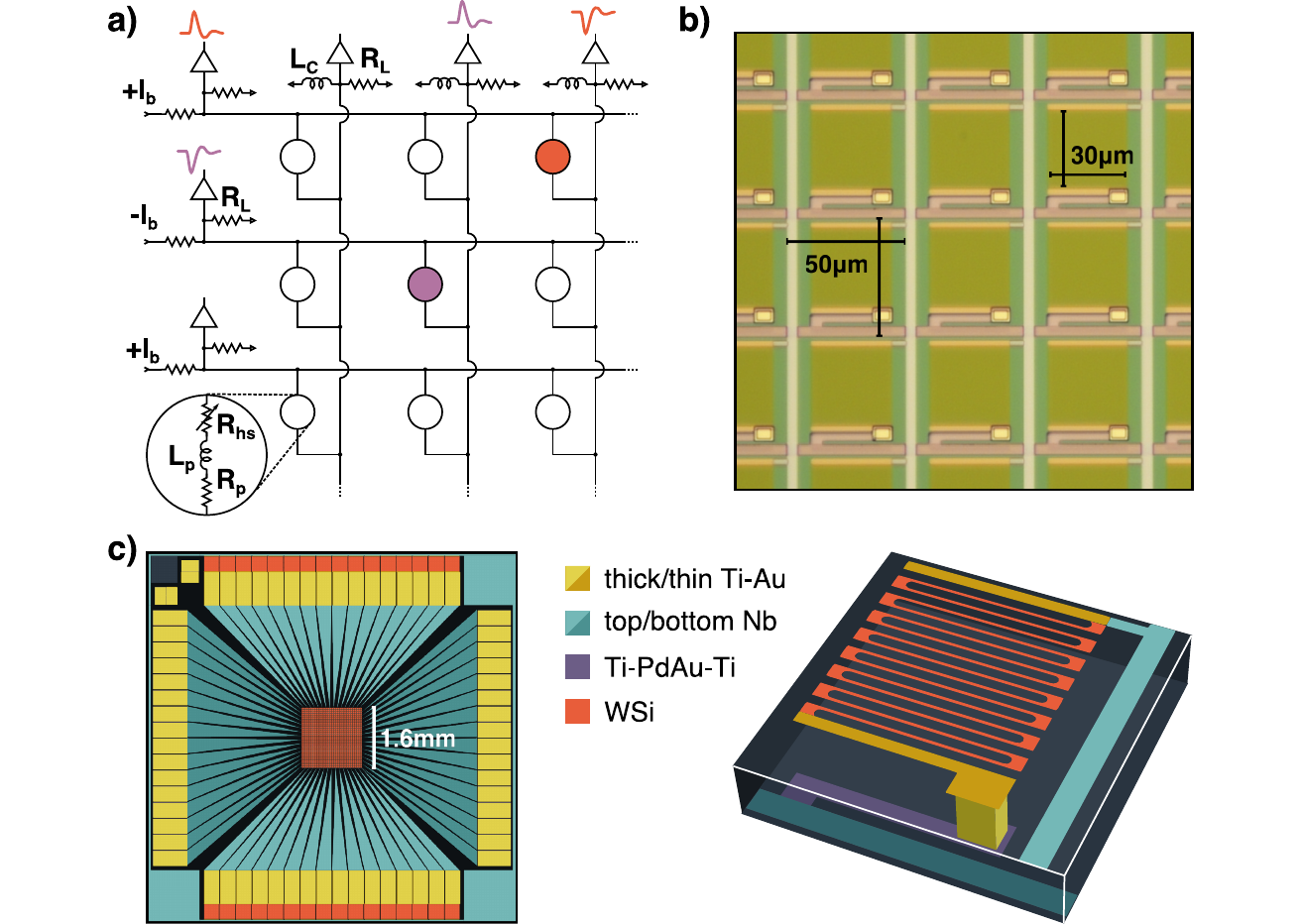}
\caption{a) Schematic of the row-column array. b) Optical micrograph of the fabricated array showing the pixel pitch and size. c) Layout of the array. Left: Chip-scale design showing the Nb leads (teal), Au bond pads (yellow), and WSi column inductors (red). Right: 3D cartoon of a unit cell of the array. The top layer containing the SNSPD and column wiring is separated from the bottom layer containing the PdAu resistor (purple) and row wiring by 250~nm of SiO\(_2\). The SNSPD meander is not shown to scale.}\label{fig:design}
\end{figure}

\section{Row-column multiplexing}\label{sec:rowcol}
The row-column multiplexing scheme has previously been described for a $2\times2$ array \cite{Verma2014} and an $8\times8$ array \cite{Allman2015}. In this scheme, $N\times N$ pixels can be read out using $2N$ lines. As shown in Fig.~\ref{fig:design}a, each pixel consists of an SNSPD and a series resistor with resistance $R_p$. One end of each pixel is connected in parallel to other pixels in the same row, and the other end of each pixel is connected in parallel to other pixels in the same column. Current is sourced to each row, distributed equally among the row's pixels by the series resistors, and sunk to ground through inductors on each column ($L_C$). Amplifiers on each column and row are used to read out photon detection events. In the specific set-up used here, DC-coupled amplifiers are used; therefore, some of the bias current also flows to ground through the $R_L = 50~\Omega$ resistors on each readout line. When a nanowire detects a photon, it develops a resistance, $R_{hs}$, on the order of 1~k$\Omega$. Current is rapidly diverted from the pixel, leading to voltage pulses of opposite polarities at its row and column readout amplifiers. Coincidences between row and column events are used to determine which of the $N^2$ pixels fired.

One downside of row-column multiplexing is the inability to image multi-photon events. For example, if two photons are detected at the same time by the pixel in row 1, column 3 (r1c3) and the pixel r2c2 (see Fig.~\ref{fig:design}a), simultaneous time tags will be recorded for rows 1 and 2 and for columns 2 and 3. It is clear that two events occurred, but the readout cannot distinguish between the pair of events being produced by r1c3 and r2c2 or by r1c2 and r2c3. In a real system with timing jitter, ambiguity can occur when two events arrive within the timing resolution of the readout, which can limit the maximum count rate of the array.

Another drawback of row-column multiplexing is current redistribution. When one pixel fires, some current is also redistributed into other pixels in the firing pixel's row and column, leading to smaller signals at the outputs and variation in a pixel's current over time. This effect can be reduced by increasing the pixel inductance ($L_p$) or series resistance ($R_p$).

\section{Fabrication}
An optical micrograph of the fabricated array is shown in Fig.~\ref{fig:design}b, and the layout of the array is shown in Fig.~\ref{fig:design}c. The array covers a $1.6 \times 1.6$ mm square with a 50~\um\ pixel pitch. The nanowire region in each pixel is $30 \times 30$~\um, resulting in a $36\%$ fill factor.

The array was fabricated at the NIST Boulder Microfabrication Facility. Fabrication begins with a 3~inch silicon wafer capped with a 150~nm-thick layer of thermally-grown SiO\(_2\). Resistors were defined at each pixel by photolithography and sputtering of a Ti / PdAu / Ti trilayer with thicknesses of 2~nm / 35~nm / 2~nm, followed by liftoff in acetone. The thickness of the PdAu layer was chosen to provide a resistance of $\sim 8 ~\Omega/\square$ at a temperature of 1~K (the approximate operating temperature of the array), yielding a total resistance at each pixel of $\sim 50~\Omega$. Although the designed resistance at each pixel was $50~\Omega$, the measured average per-pixel resistance in the fabricated array was estimated to be $\sim 200~\Omega$. This resistance was determined based on the slope of the superconducting branch of the I-V curve for a single row scaled by 32 for the number of pixels in the row. The same PdAu layer also defines the contact pads for the rows. Each row contact pad connects in parallel to a single row of 32 resistors through a superconducting Nb wiring layer which is 50~nm thick. This layer is defined by photolithography, sputtering of the Nb layer, and liftoff. A 250~nm-thick SiO\(_2\) dielectric layer is then deposited by plasma enhanced chemical vapor deposition (PECVD). Vias through the SiO\(_2\) dielectric are defined by photolithography and reactive ion etching in a CHF\(_3\)/O\(_2\) plasma. These vias connect the SNSPD to the underlying resistor. Vias are also etched over the row contact pads in the same step. The vias are filled with 5~nm Ti / 350~nm Au by photolithography, electron beam evaporation, and liftoff. A 50~nm-thick layer of Au is then patterned and deposited using electron beam evaporation and liftoff which serves to connect the SNSPD to the via. 

The 3~nm-thick WSi layer which is used to fabricate the SNSPD is deposited by cosputtering from separate tungsten and silicon targets at room temperature, followed by a 2~nm-thick amorphous Si cap which serves to protect the WSi from oxidation and subsequent processing steps. A square of WSi having a width and length of 34~\um\ is then defined across the 50~nm-thick gold contacts at the location of each pixel by photolithography and RIE etching in an SF\(_6\) plasma. The same WSi layer is also used to fabricate the column inductors. The wire composing the inductor is 1~\um\ wide, and has a total length of 74~mm. Electron beam lithography is then performed to define the nanowires using PMMA resist. The nanowires have a width of 180~nm, and the pitch of the meandering wire pattern is 260~nm. The PMMA pattern is transferred into the underlying WSi layer using RIE etching in SF\(_6\). The column wires connecting columns of 32 SNSPDs are patterned by sputtering of 50~nm of Nb and liftoff. The Nb wiring layer connects directly to the gold contact pad at each SNSPD in a column. Each Nb wire terminates at the WSi column inductor where the two layers are again connected by a 50~nm-thick gold contact pad. The Nb layer also defines the ground plane on the opposite end of the inductor. After patterning, the wafers are diced into 1~cm dies for mounting into standard plastic-mounted chip carriers (PLCCs), enabling integration with the existing 64-channel readout electronics.

\section{Readout and control electronics}
The 64-channel time-tagging readout electronics were developed as part of the ground receiver for NASA's Deep Space Optical Communications project \cite{Biswas2018}. The first stage of amplification is provided by DC-coupled amplifiers with a $50~\Omega$ input impedance located at 40~K, with further amplification at room temperature. A 64-channel time-to-digital converter (DotFast/UQDevices model TDM64-800 \cite{Dotfast}) converts the SNSPD pulses to time tags with a timing resolution of 15.625~ps and FWHM timing jitter below 50~ps. The TDC has a built-in comparator front end capable of triggering on the rising or falling edge of positive or negative signals. Time tags are output over PCI~Express (PCIe) at rates of up to 900~MTag/s. Currently, the tags are written to a file for post-processing, but the PCIe output is compatible with real-time analysis using an FPGA.

\begin{figure}[t]
\centering\includegraphics{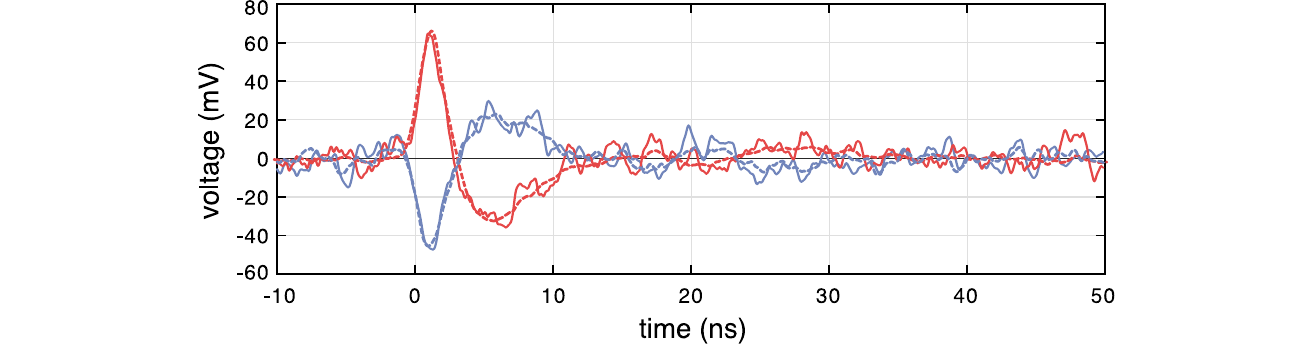}
\caption{Examples of positive (red) and negative (blue) pulses from one column of the array measured at the TDC input. The array was biased at 3~\uA\ per pixel. Solid lines represent individual pulses and dashed lines are time-averaged pulses. The narrow pulse shape and overshoot are caused by high-pass filtering by the cryogenic amplifiers.}\label{fig:osc}
\end{figure}

The rows of the array are biased using four NI-9264 analog voltage output modules, and the bias current is introduced through 10~k$\Omega$ resistors on the 40~K amplifier board. We found that the array became unstable when all rows were biased at the same voltage. The instability is most likely due to the bias current from all detectors in a column approaching or exceeding the switching current of the column's inductor. The inductor's switching current was measured to be approximately 90~\uA, while the total current produced in each column for the measurements described here reached 140~\uA. This instability disappeared when alternating positive and negative voltages were applied to alternating rows, nulling the current through the column inductors. The array was thus biased with a positive voltage applied to all even rows and a negative voltage applied to all odd rows. Because the comparator threshold can either be positive or negative, for each measurement, time tags were first acquired while triggering on negative pulses from the odd rows and positive pulses from all columns and then a second acquisition was taken while triggering on positive pulses from the even rows and negative pulses from all columns. Example positive and negative pulses from the same column are shown in Fig.~\ref{fig:osc}. To make an operational array beyond this proof of concept, the column inductors can either be eliminated or be designed to have a higher switching current. The column inductors allow for readout using AC-coupled amplifiers but are not necessary with the DC-coupled readout scheme used here.

Once the time tag files are acquired, we can assign row-column time tag pairs to their corresponding pixels. To do this, time tags are read out of the file, and sequential pairs of time tags that are separated by less than a specified coincidence window are recorded as events for the pixel from the corresponding row and column. Due to varying delays between the channels on chip, in the amplifier boards, and in the TDC's FPGA, each pixel has a different delay between the time when its row registers an event and the time when its column registers an event. Without calibration, all row-column time differences fall within a 6~ns range; we thus use 6~ns as the coincidence window unless otherwise stated. By calibrating the row and column delays, it is also possible to adjust time stamps by their channel's delay and null out the average row-column time difference. The DotFast TDC has the capability to compensate for time delays between its channels before outputting sorted time tags.

\section{Measurements}
The array is operated at a temperature of 730~mK on a He-3 sorption cooler and is illuminated with 1550~nm light using free-space optics. A set of three short-pass filters mounted at the 40~K and 4~K radiation shields blocks room-temperature blackbody radiation while maintaining high transmission at 1550~nm. Optical losses in the filters are estimated to be less than 4\%.

\begin{figure}[t]
\centering\includegraphics{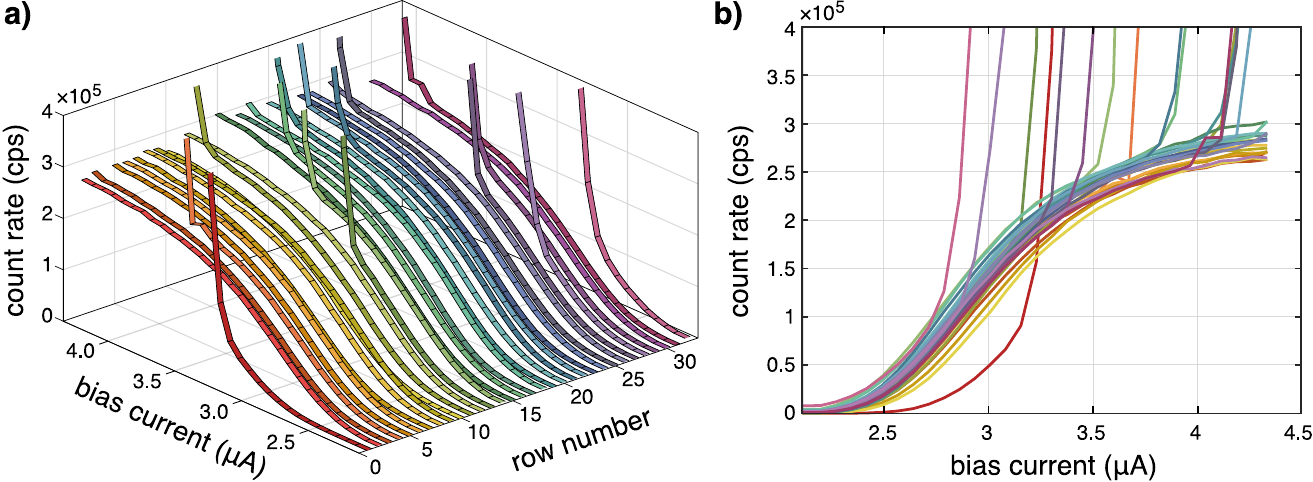}
\caption{Count rate vs. average pixel bias current summed across each of the 32 rows under 1550~nm flood illumination. a) 3-D view of the count rate showing the spatial distribution of efficiency and hot pixels across the array. In general, the hot pixels appear to be evenly distributed spatially, although the two edge rows do contain some of the most constricted pixels. b) 2-D view of the same data in (a). A clear inflection point is visible in the count rate vs. bias curve. Above 4.3~\uA, more than half of the array's rows are affected by hot pixels, although the overall percentage of hot pixels is much lower.}\label{fig:pcr}
\end{figure}

\subsection{Bias-dependent photoresponse and yield}
Fig.~\ref{fig:pcr} shows the total count rate vs. bias current curves for each of the 32 rows under 1550~nm flood illumination. The bias current is calculated as the average current in each pixel assuming uniform current distribution. Four pixels out of the 1024 pixels in the array show no photoresponse at any bias, and two pixels show a much lower photoresponse than their neighbors, corresponding to a baseline yield of 99.4\%. As the bias current increases, the array develops more ``hot'' pixels with elevated count rates, which also decrease the yield of functional pixels. The hot pixel behavior is due to relaxation oscillations in the presence of the DC-coupled readout as the pixel's bias current surpasses its switching current. The switching current could be suppressed for these pixels because of variations in the nanowire thickness or width across the array, variations in the pixel resistor values, or constrictions in the nanowire. Because these hot pixels tend to have similar count rates to their neighbors at lower bias points, constrictions are the most likely cause. As the bias current increases, some of these hot pixels eventually switch to their normal state.

As the rows approach saturated internal efficiency, there remains about 10\% variation in count rate across the rows. The pixel-to-pixel variation can arise from 1) variation in illumination, 2) variation in bias current due to differences in the on-chip pixel resistors, 3) variation in bias current due to the amplifier board, 4) variation in the comparator's triggering efficiency due to differences in amplifier noise or gain across the channels or 5) variations in the nanowire fabrication.

\subsection{Detection efficiency and imaging capabilities}
\begin{figure}[t]
\centering\includegraphics{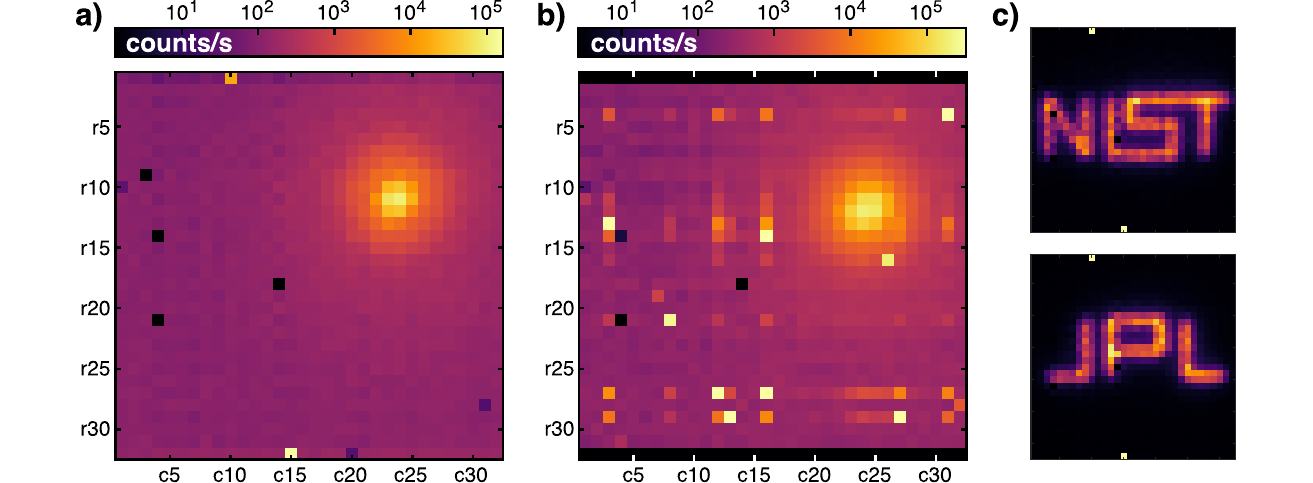}
\caption{Imaging capabilities of the array. a, b) Log-scale count rate across the array under illumination with a focused laser beam. a) Approximately 3~\uA~bias current, adjusted to equalize background counts across the rows. The laser spot photon flux is $8.5\times10^6$~photons/s (ph/s), and the color scale is clipped at count rates above $1.6\times10^5$~counts/s (cps).  b) 4~\uA~bias current, with rows 1 and 32 biased at 0~\uA. The laser spot photon flux is 14.9~Mph/s, and the scale is clipped at count rates above $3.3\times10^5$~cps. At the lower bias, there are fewer hot or dead pixels, and no cross-talk is evident. At the higher bias current, misattributions due to hot pixels are evident in streaks and false hot pixels. The hot pixel with the highest count rate (r13c3) has a count rate of 4.6~Mcps. c) Persistence images as a laser spot is swept across the array to spell out text using a steering mirror and a voltage-controlled attenuator (see Visualization 1). The array bias current is 3~\uA.}\label{fig:spot}
\end{figure}

 To measure the system detection efficiency of the array, we imaged a focused laser beam with a known power. A 2-D Gaussian fit to the resulting count rate across the array yielded a pixel efficiency of $12\%$ at a bias current of 3~\uA, and $23\%$ at a bias current of 4~\uA. Combined with the $36\%$ fill factor of the array, the total system detection efficiency is calculated to be $4\%$ at 3~\uA\ and $8\%$ at 4~\uA. In the future, the efficiency of the array can be enhanced significantly by embedding the pixels in an optical cavity and optimizing the fill factor. Figs.~\ref{fig:spot}a and b show the log-scale count rate of the array under illumination by the laser spot. The image in Fig.~\ref{fig:spot}a was taken at a bias current of approximately 3~\uA\ with a 1.2~ns coincidence window, and the image in Fig.~\ref{fig:spot}b was taken at a bias current of approximately 4~\uA\ with a 0.6~ns coincidence window. The coincidence windows were chosen to collect > 99.5\% of counts after a timing calibration was applied to the rows and columns, based on the measured jitter for each bias current, as discussed in Section~\ref{sec:jitter}. At the lower bias current, five dead pixels and two hot pixels are evident. At the higher bias current, there are nine more hot pixels (count rate > 100~kcps). As seen in Fig.~\ref{fig:spot}b, the elevated count rates in these pixels' rows and columns lead to misattribution of an event's originating pixel, with false hot pixels and streaks appearing at the intersection of rows and columns containing elevated counts originating from either hot pixels or the laser spot. The false pixels are clearly visible in this log-scale image; however, misattributions consist of at most 0.6\% of the corresponding hot pixels' counts. At the low bias current, no misattribution is evident.

As a demonstration of time-dependent imaging using the array, we swept a laser spot across the array to spell out text. Persistence images of counts across the array are shown in Fig.~\ref{fig:spot}c, and movies are available online (see Visualization 1). For these measurements, the array was biased at 3~\uA.

\begin{figure}[t]
\centering\includegraphics{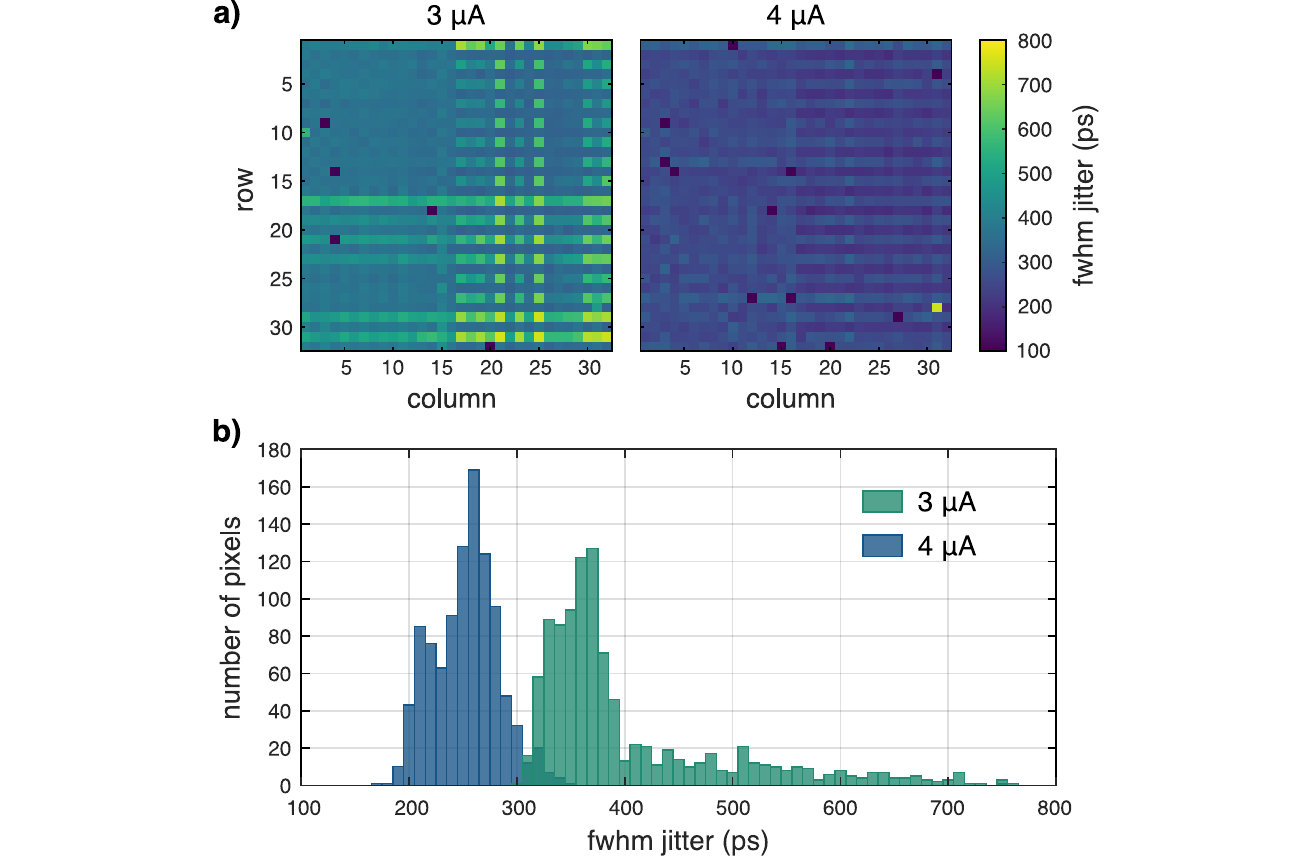}
\caption{Jitter measurements. a) Maps of the FWHM jitter measured for each pixel of the array at bias currents of 3~\uA\ (left) and 4~\uA\ (right). b) Histograms of the data in (a) showing the distribution of jitter across the array measured with the array biased at 3~\uA\ (green) and at 4~\uA\ (blue). At the lower bias current, the average pixel jitter was 400~ps, but the jitter distribution is peaked near 365~ps with a long tail towards higher jitter values. At the higher bias current, the average pixel jitter was 250~ps.}\label{fig:jitter}
\end{figure}

\subsection{Jitter}\label{sec:jitter}
Jitter measurements were performed using a mode-locked laser to flood-illuminate the array. Pixel time tags were recorded as the average of the corresponding row and column time tags, and a 10~ns coincidence window was used in the time tag analysis to avoid any artificial reduction of the jitter. The resulting timing histograms were fitted with a Gaussian distribution to determine the full-width half-maximum (FWHM) timing jitter. For some pixels, it was not possible to obtain a good fit, due either to lack of counts or interference from neighboring hot pixels. Fig.~\ref{fig:jitter}a shows the resulting maps of the FWHM jitter measured for each pixel of the array at bias currents of 3~\uA\ and 4~\uA, and Fig.~\ref{fig:jitter}b shows the same data binned into histograms. The measurements were taken at count rates of approximately 5~kcps per pixel. At the lower bias current, the average jitter was 400~ps, and at the higher bias current, the average jitter was 250~ps. These values are much higher than the <~100~ps jitter of a typical SNSPD and the <~25~ps jitter of the TDC. The high jitter can be attributed both to the pixels' relatively low bias currents and to current redistribution in the array, as described in Section~\ref{sec:rowcol}. The current redistribution diverts signal away from the amplifier chain where it can be detected and into other pixels. One effect of this redistribution is to decrease the signal level of the voltage pulse and to make the jitter more susceptible to amplifier noise. For example, in Fig.~\ref{fig:jitter}a, the jitter map for the low bias current shows higher jitter for certain rows and columns, indicating that these channels have lower gain or higher noise. Another effect of current redistribution is to produce fluctuations in the bias current through each pixel, which then add to jitter through variations in pulse height. Future row-column arrays can minimize these effects with further optimization of the readout to decrease the amplifier noise or the pixel inductance and resistance to improve the signal. The use of a constant fraction discriminator instead of a fixed-threshold comparator can also reduce the jitter that originates from temporal walk associated with varying pulse heights.

As described in Section~\ref{sec:rowcol}, the array's jitter is relevant for avoiding misattribution. If the differential timing delay for each readout channel is calibrated out, then the minimum coincidence window will be limited by jitter. The smaller the coincidence window, the less likely that two events will occur during the coincidence window.

To probe the array's maximum count rate, we biased either half or a quarter of the rows of the array at 3~\uA\ and applied an increasing optical flux. The whole array was not biased at once in order to avoid the TDC's maximum count rate of 900~MTags/s. With the efficiency calculated from the totalized count rate across the 8 or 16 biased rows instead of from coincidences, the efficiency showed no degradation up to count rate of 10~Mcps per row and an approximately 10\% reduction at 30~Mcps per row. This calculation neglects the effects of misattribution on the maximum count rate. For example, if we assume that the photon arrival times follow a Poissonian distribution, and we use a coincidence window of 6~ns, at an average count rate of 320~Mcps across the whole array, approximately 60\% of detected photons will arrive within the coincidence window of another photon, leading to misattribution errors for 30\% of the detected events. For a coincidence window of 1.2~ns, misattribution errors decrease to approximately 3\%. However, at higher count rates, there is likely to be more current redistribution through the array and thus higher jitter.

\section{Conclusion}
In summary, we have fabricated an SNSPD array with 1024 pixels and a $1.6 \times 1.6$~mm area with a row-column multiplexing architecture. Using a 64-channel time tagging readout, we found that the array has a baseline yield of over 99\%, a system detection efficiency of up to 8\% at 1550~nm, and jitter of 250 - 400~ps. Our results show that it is feasible to extend the previously-demonstrated row-column multiplexing architecture to kilopixel arrays and to yield SNSPDs over millimeter-scale active areas.

\section*{Acknowledgments}
Part of this research was performed at the Jet Propulsion Laboratory, California Institute of Technology, under contract with the National Aeronautics and Space Administration. This work was supported in part by the DARPA DETECT program and Center Innovation Funds (CIF) from NASA STMD.

%%%%%%%%%%%%%%%%%%%%%%% References %%%%%%%%%%%%%%%%%%%%%%%%%

\end{document}